# Zeolitic imidazolate framework-coated acoustic sensors for room temperature detection of carbon dioxide and methane

Jagannath Devkota[1,2,§], Ki-Joong Kim[1,2], Paul R. Ohodnicki[1,3,§], Jeffrey T. Culp[1,2], David W. Greve[4,5], and Jonathan W. Lekse[1]

[1]National Energy Technology Laboratory, Pittsburgh, PA 15236

[2]AECOM Pittsburgh, PA 15236

[3]Department of Materials Science and Engineering, Carnegie Mellon University, Pittsburgh, PA 15213

[4]Department of Electrical and Computer Engineering, Carnegie Mellon University, Pittsburgh, PA 15213

[5]DWGreve Consulting, Sedona, AZ 86336

[§] Corresponding authors: jagannath.devkota@netl.doe.gov; paul.ohodnicki@netl.doe.gov

## Abstract

Integration of nanoporous materials such as metal organic frameworks (MOFs) with sensitive transducers can result robust sensing platforms for monitoring gases and chemical vapors for a range of applications. Here, we report on an integration of the zeolitic imidazolate framework – 8 (ZIF-8) MOF with surface acoustic wave (SAW) and thickness shear mode quartz crystal microbalance (QCM) devices to monitor carbon dioxide ($CO_2$) and methane ($CH_4$) at ambient conditions. The MOF was directly coated on the custom fabricated Y-Z $LiNbO_3$ SAW delay lines (operating frequency, $f_0$ = 436 MHz) and AT-cut Quartz TSM resonators (resonant frequency, $f_0$ = 9 MHz) and the devices were tested for various gases in $N_2$ at ambient condition. The devices were able to detect the changes in $CO_2$ or $CH_4$ concentrations with relatively higher sensitivity to $CO_2$, which was due to its higher adsorption potential and heavier molecular weight. The sensors showed full reversibility and repeatability which were attributed to the physisorption of the gases into the MOF and high stability of the devices. Both types of the sensors showed linear responses relative to changes in the binary gas compositions thereby allowing to construct calibration curves which correlated well with the expected mass changes in the sorbent layer based on mixed-gas gravimetric adsorption isotherms measured on bulk samples. For 200 nm thick films, the SAW sensitivity to $CO_2$ and $CH_4$ were $1.44\times10^{-6}$/vol-% and $8\times10^{-8}$/vol-%, respectively against the QCM sensitivities $0.24\times10^{-6}$/vol-% and $1\times10^{-8}$/vol-%, respectively which were evaluated as the fractional change in the signal. The SAW sensors were also evaluated for 100 nm – 300 nm thick films, the sensitivities of which were found to increase with the thickness due to the increased number of pores for adsorption of larger amount of gases. In addition, the MOF-coated SAW delay lines had a good response in wireless mode, demonstrating its potential to operate remotely for detection of the gases at emission sites across the energy infrastructure.

**Graphical Abstract:**

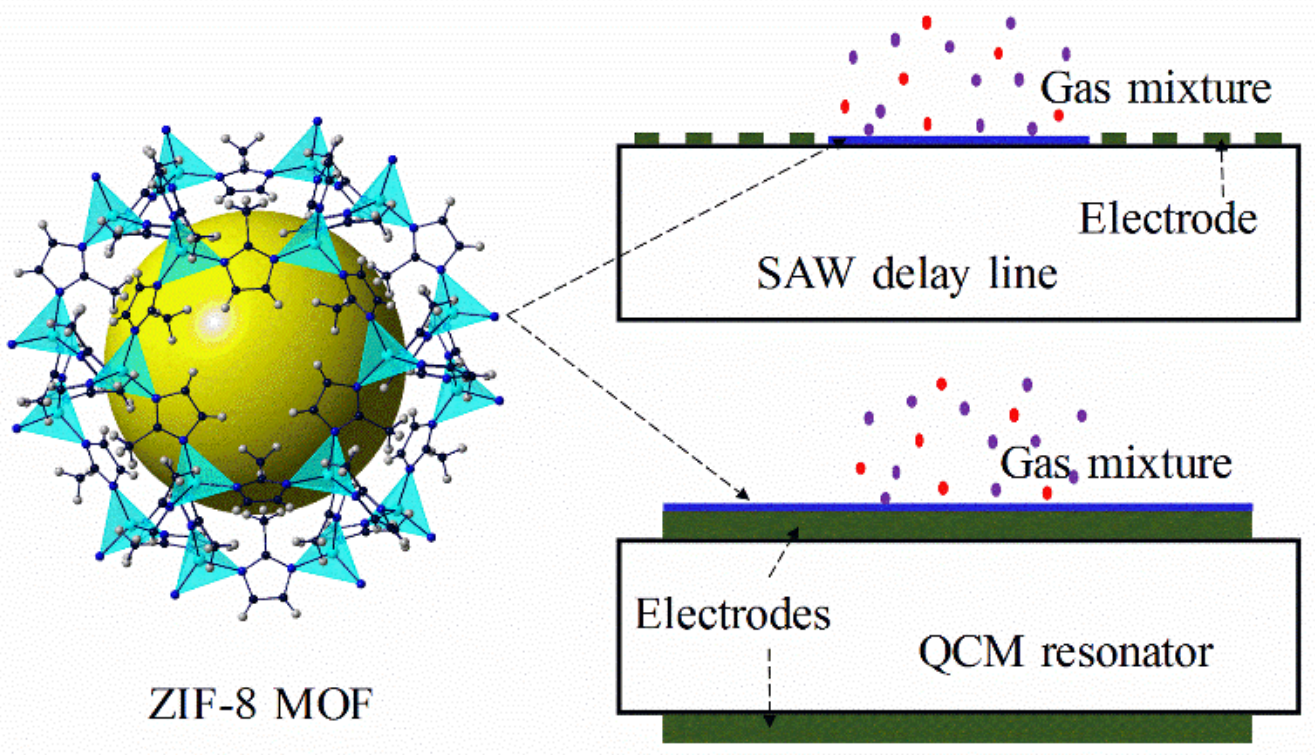

## 1. Introduction

Greenhouse gas emissions such as carbon dioxide ($CO_2$) and methane ($CH_4$) from the processing and use of fossil fuels create environmental concerns and challenges in fuel transportation [1]. Even though sensors are available for detection and analysis of these gases, there exist opportunities for enhanced monitoring capability of geographically distributed infrastructure such as through advanced devices that are sufficiently low cost for ubiquitous deployment and can be interrogated wirelessly. Examples include $CH_4$ leak detection for the natural gas and oil infrastructure (e.g. pipelines, compressors, active and abandoned wells), $CO_2$ monitoring for understanding plume migration within and near geological formations in carbon sequestration applications, and detection of such species in wellbore integrity monitoring and drilling applications amongst others. Therefore, a need exists for developing novel, cost-effective, and reliable sensors with high sensitivity and selectivity as well as ability to operate wirelessly for monitoring these gases. Surface acoustic wave (SAW) devices, which are highly sensitive, low cost, small, and operational in wireless and passive modes, can satisfy all requirements through selective detection of chemical species at ambient conditions when coated with proper sensing materials [2]. In these sensors, the coating materials play a key role in interacting with the target gases and coupling the interaction with the propagating wave characteristics. Their properties not only determine the selectivity, reversibility, and repeatability but also can affect the sensitivity and response kinetics [2]. Variety of materials such as metal oxides [3, 4], carbon nanotubes [5], graphene-based composites [6], and polymers [7-10] have been applied on SAW devices for gas detection. However, the majority of these materials have either poor interaction with gases at ambient conditions or poor selectivity thereby limiting their applications. For instance, the sensors based on metal oxides can possess high sensitivity and can be engineered for adequate selectivity, but they need elevated temperatures for interaction with gases whereas those based on polymers have limited sensitivity and selectivity. Inert nature of $CH_4$ adds extra complication in developing a robust sensor for its monitoring even in emission sites. To address these issues, there is a need of (i) identifying novel materials that can incorporate large amount of specific gases by adsorption at ambient conditions and (ii) integrating them with sensitive transducers like SAW devices.

In recent years, a novel class of nanoporous crystalline materials composed of metallic ions and organic linkers, the metal-organic frameworks (MOFs), have attracted considerable attention in sensing and other applications due to their diverse structure with uniform pores, large surface area, tunable gas adsorption properties at ambient conditions as well as at high pressures, and good thermal stability [11-14]. They can physisorb large amount of gases into their pores whose structure, aperture, and sizes can be engineered to tune the adsorption for desired selectivity and sensitivity [11, 15].They have been extensively studied for high pressure applications including gas storage, separation, and heterogeneous catalysis[16, 17]. However, their use as sensing materials is still in their infancy due to challenges to integrate them with electronic devices including acoustic transducers[13, 15, 18-24]. Zeitler *et. al.* reported a computational study to investigate a possibility of incorporating a range of MOFs with QCM, SAW, and microcantilever devices for $CH_4$ detection with high sensitivity [25]. They pointed out that the performance of SAW sensors coated with MOFs can have better sensitivity than that of QCM sensors. Yamagiwa *et al.* integrated $Cu_3(BTC)_2$ and $Zn_4O(BDC)_3$ with QCM resonators for detection of humidity and volatile organic compounds [26]. Similarly, some other research groups also applied $Cu_3(BTC)_2$ on QCM resonators to detect humidity and some organic analytes [20, 27]. Robinson *et al.* integrated $Cu_3(BTC)_2$ with SAW devices to develop humidity sensors [14]. More recently, Paschke *et al.* applied MFU-4 and MFU-4*l* MOFs on SAW devices and investigated their gas uptake kinetics [21]. Besides a few of these, thousands of other MOFs reported to date (and hundreds of thousands more MOFs conceivable) display a wide range of gas adsorption behaviors and thereby open the possibility of tailoring adsorption-based sensors to a particular analyte and sensor arrays for compositional analysis of multicomponent mixtures.

Although there are some computational approaches [28] to screen MOFs as sensing layers, experimental investigations are critical for determining appropriate methods of fabricating dense thin films on a sensor surface and verifying that the gas adsorption properties of the bulk material are preserved when incorporated into a sensing device. Here, we investigate the potential of ZIF-8 as a sensing layer on SAW and QCM devices for detection of various concentrations of $CO_2$ and $CH_4$ in $N_2$ at room temperature and atmospheric pressure. It is a non-conducting crystalline (space group: *I43m*) MOF composed of zinc ions and 2-methylimidazole organic linkers (**Figure 1(a)**) [29, 30]. This particular MOF was chosen as it is well characterized for its structural as well gas adsorption properties and can be grown as uniform thin films on various substrates by simple methods [19, 29, 31]. As such, ZIF-8 provides an excellent model material to evaluate how effectively the gas adsorption properties of the bulk material can be transferred to a functional device. Its structure contains large cages (diameter 11.6 Å) with narrow apertures (width 3.4 Å) formed by six-membered rings [19]. While the crystalline pore aperture is smaller than the kinetic diameters of $N_2$ and $CH_4$, the ligands which frame the pore opening are able to move slightly and increase the effective pore aperture. Due to this ligand flexibility, the sieving aperture in ZIF-8 is approximately 6 Å, which is more than large enough for the adsorption of small gases such as $CO_2$, $CH_4$, and $N_2$ through physical processes [32]. While this MOF is known to have only moderate gas adsorption selectivity; low water affinity and the good stability at ambient conditions of the material make it a good candidate for niche applications in humid environments such as leak monitoring along natural gas pipelines where competitive adsorption of multiple gases is minimized and only the detection of a particular analyte gas in an air background is needed [33-35]. Recently, a terahertz spectroscopy-based sensor was developed using this MOF for detection of $CO_2$ and some other chemical vapors at ambient conditions [36]. In the current study, we investigated systematically the possibility of using this MOF to develop mass-based SAW and QCM sensors to monitor $CO_2$ and $CH_4$ as well as other analytes in a background of $N_2$. The developed sensors gave linear responses towards linear changes in $CO_2$ and $CH_4$ concentrations with a little or no measurable response to competing gases such as CO and air. As expected, the mass-based sensor response showed a larger sensitivity towards $CO_2$ compared to $CH_4$ due to the additive effects of higher adsorption affinity of ZIF-8 for $CO_2$ over $CH_4$ and the higher molecular weight of $CO_2$. While both types of sensors were able to monitor various concentrations of $CO_2$ and $CH_4$, SAW sensor showed greater sensitivity in terms of the fractional change in output signal. Furthermore, we showed the wireless gas detection ability of the MOF-coated SAW devices. The mass of the gases adsorbed into a unit volume of ZIF-8 in the sensor films as determined from the sensor responses agreed with the amounts adsorbed in a bulk sample of ZIF-8 powder obtained by gravimetric methods for both pure gases and $CO_2/N_2$ and $CH_4/N_2$ mixtures. In addition, the sensitivities of the SAW sensors were evaluated for varied thickness of the film (100 – 300 nm) and were found to increase with the film thickness which was explained based on the increased pore numbers into the thicker films for adsorption of larger amount of gas molecules. A summary of the sensors' working principles is given below:

***Working Principle:***

SAWs propagate along the surface of a material and decay exponentially along the depth whereas the bulk acoustic waves (BAWs) propagate across the material's body (thickness shear mode propagation). The gas sensors based on these waves rely on changes in their velocity (frequency or phase) or amplitude induced by the changes in mass, conductivity, or stiffness of the overlayer upon interaction with gases [2]. In the current study, we monitor the change in the phase velocity caused by the gas adsorption in the ZIF-8 films applied on to the SAW and BAW transducers (**Figures 1(b)** and **1(c)**, respectively). The velocity is usually the preferred parameter in experiments as it remains unaffected from electromagnetic interferences unlike the amplitude. When the coated ZIF-8 films incorporate gas molecules (which happens by physisorption [37] in this case), there can be changes in their mass density and mechanical properties which induce a change in the wave velocity. For acoustically thin (*i.e.* $h << \lambda$, where $h$ is the film thickness and $\lambda$ is the wavelength) films of soft materials, the net mass change is much higher than the elastic changes due to interaction with gases so that the velocity change, $\Delta v \propto -\Delta m$ (where $\Delta m$ is the loaded mass to the films) to the first order [38-40]. For SAW sensors coated with acoustically thin films of non-conducting, lossless, and isotropic materials, the velocity change upon gas exposure can be written as [39, 41],

$$\frac{\Delta v}{v_0} = -c_m f_0 \left(\frac{\Delta m}{A}\right) \qquad (1)$$

where $c_m$ is the mass sensitivity coefficient of the substrate ($c_m = 5.505 \times 10^{-7}$ $cm^2s/g$ for Y-Z $LiNbO_3$) and $A$ is its piezoelectrically active area. For a full coverage of the piezoelectrically active area, one can write,

$$\frac{\Delta v}{v_0} = \frac{\Delta f}{f_0} = -\frac{\Delta \phi}{\phi_0} \qquad (2)$$

where $v_0$, $f_0$ and $\phi_0$ ($= 2\pi f_0 t_0$, with $t_0$ being the time delay in the reference state) are the velocity, frequency, and phase, respectively in the reference state. A correction factor can be introduced in Eq. (2) if the piezoelectrically active area is partly covered by the sensing overlay [42]. For bulk wave-based QCM sensors coated with thin films (so that the change in resonance frequency is less than 2%), the mass loading effect to the wave velocity is given by the Sauerbrey equation [27, 39],

$$\frac{\Delta f}{f_0} = -\frac{2}{\sqrt{\rho\mu}} f_0 \left(\frac{\Delta m}{A}\right) \qquad (3)$$

where $\rho$ and $\mu$ are the density and shear modulus of the substrates (For AT-Quartz, $\rho = 2.648\ \mathrm{gcm^{-3}}$ and $\mu = 2.947 \times 10^{11}\mathrm{gcm^{-1}s^{-2}}$).

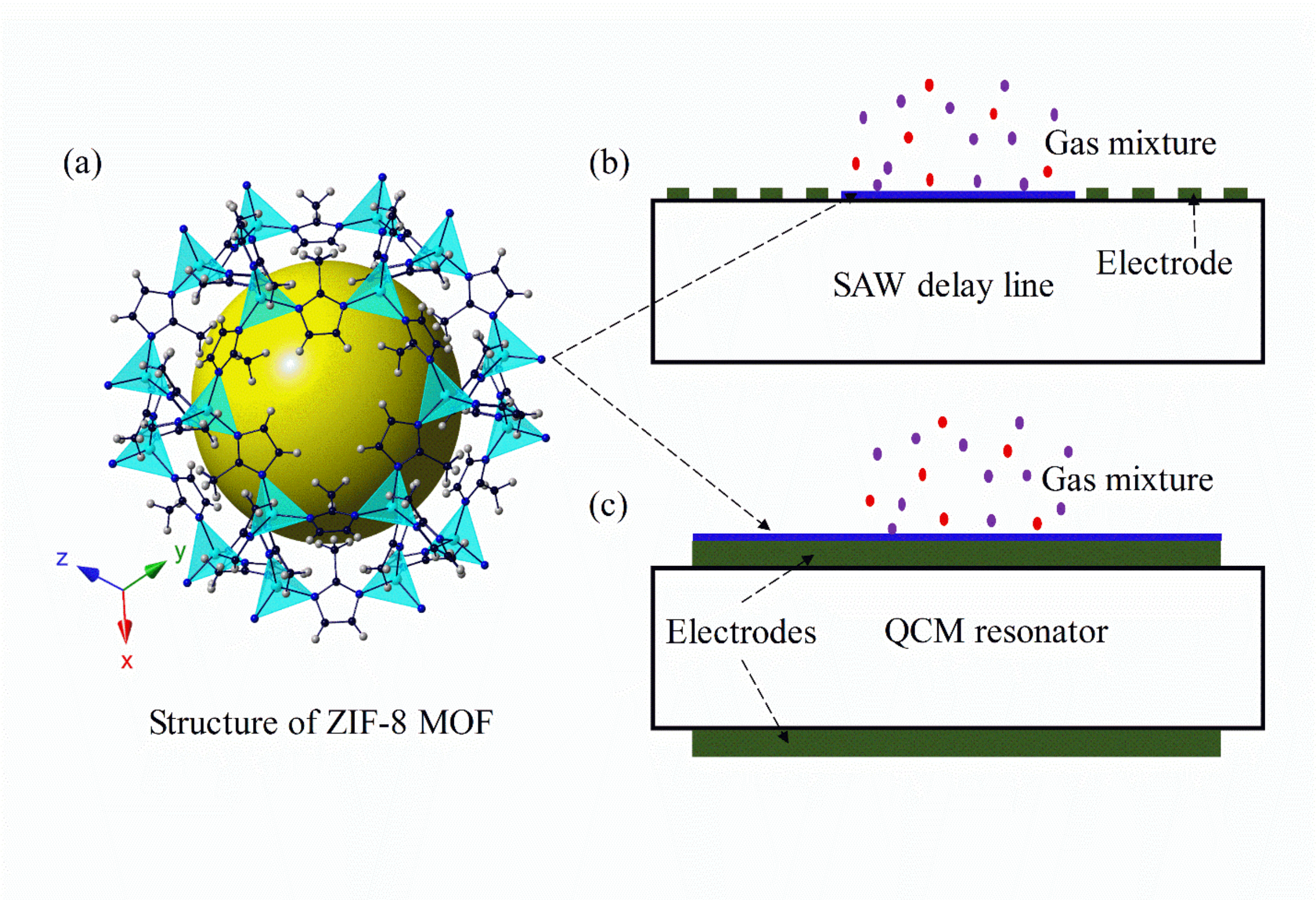

*Figure 1. Schematic of (a) a ZIF-8 crystal and its films applied to (b) surface acoustic wave (SAW) and (c) quartz crystal microbalance (QCM) transducers.*

## 2. Finite Element Analysis of SAW devices

We performed 2D finite element modeling (FEM) [43] using COMSOL 5.3 to compute the effect of ZIF-8 layer on SAW velocity in a Y-Z $LiNbO_3$ transducer for (i) no gas exposed and (ii) pure $CO_2$ and $CH_4$ exposed conditions. An eigenmode analysis was performed in plane strain mode using the geometry shown in **Figure 2(a)** for which the boundary conditions are summarized in **Table 1**. The width of the geometry was chosen equivalent to one wavelength ($\lambda = 8\,\mu m$) to reduce the computation time and periodic boundary conditions were employed to simulate its infinite length along the wave propagation. The thickness of the substrate was chosen to be five times the wavelength which should be good approximation for modeling Rayleigh wave-based SAW devices as the waves are known to be confined close to the surface [2]. The modeling parameters for Y-Z $LiNbO_3$ piezoelectric substrate were taken from COMSOL's inbuilt materials library and those for ZIF-8 layer were taken from literature [31, 44]. The density, relative permittivity, Young's modulus, and Poisson's ratio of ZIF-8 were considered as 0.95 g/cm$^3$, 2.0, 3.5 GPa, and 0.43, respectively and the films were considered lossless and isotropic. For a periodic structure with 8 µm width, the SAW resonant frequency was found to be ~436.27 MHz for a free surface (no ZIF-8 layer), which is consistent with the reported SAW velocity 3488 m/s in the substrate [39]. **Figure 2(b)** shows the surface displacements obtained at the resonance during SAW propagation on free surface case. When a 200-nm thick ZIF-8 layer was applied on the Y-Z $LiNbO_3$ surface, the velocity was predicted to decrease by 0.49%. The responses of the sensors due to the mass loaded into the 200-nm thick ZIF-8 films were approximated introducing a thermodynamic parameter, the partition coefficient, $K = C_s/C_v$, where $C_s$ and $C_v$ are the concentrations of the gases in into stationary (MOF) and volatile phases, respectively [45]. First, we evaluated the dimensionless Henry's constant (which can be considered as the partition coefficient [46]) for pure

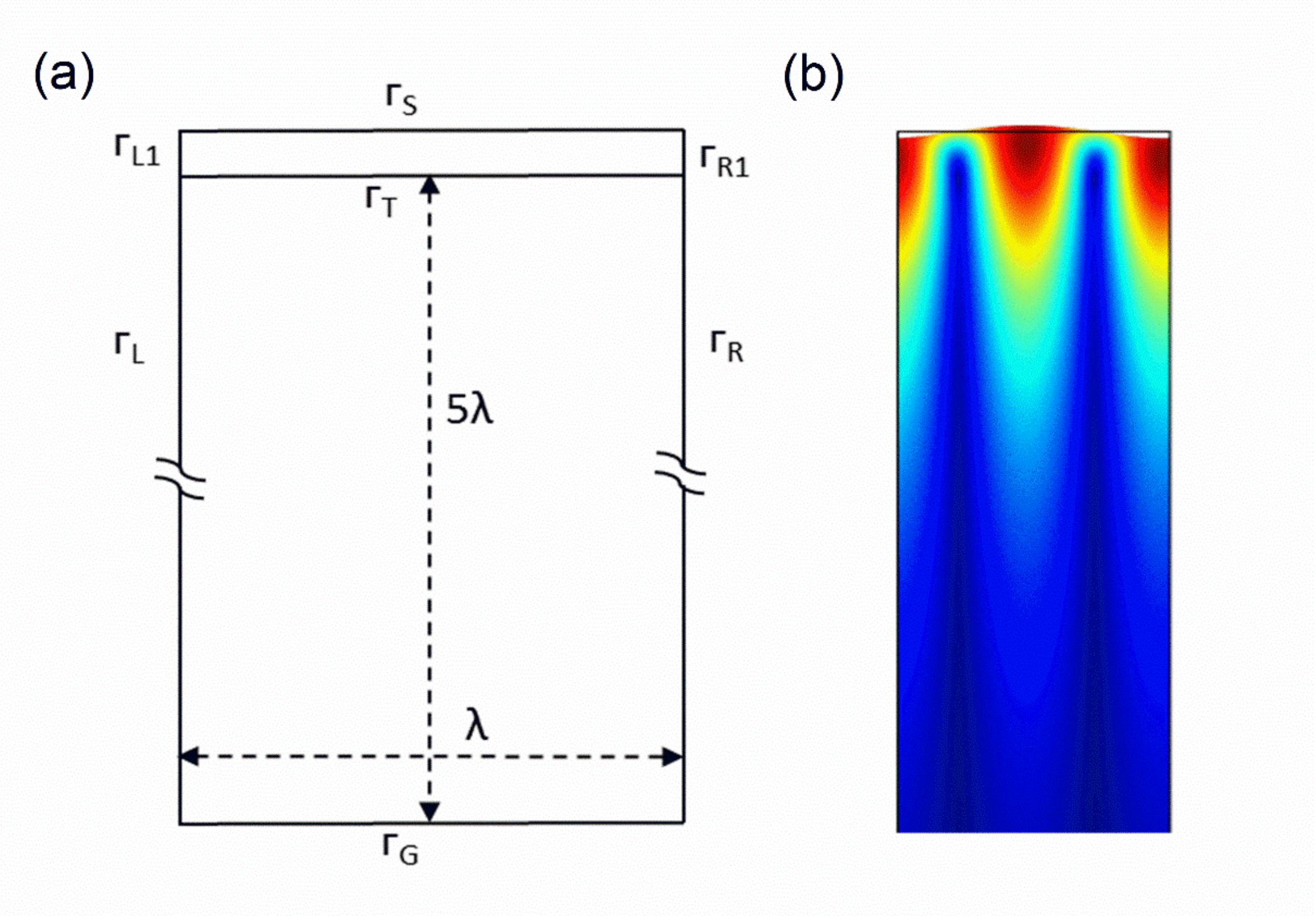

*Figure 2. (a) Geometry considered for finite element modeling of SAW devices coated with ZIF-8 film and (b) the total surface displacements on a free surface of Y-Z $LiNbO_3$ SAW device at resonance.*

$CO_2$ and $CH_4$ in ZIF-8 from the gas adsorption isotherms measured by gravimetric method at 25°C (to be discussed below) and employed it to calculate the change in mass density ($\Delta\rho = KC_v$) of the ZIF-8 films upon gas adsorption and hence to estimate the sensor response (Eq. 1) at 1 atm. The fractional decrease in the velocity of the SAW device with 200-nm thick ZIF-8 layer was estimated to be $189.3\times10^{-6}$ and $25.6\times10^{-6}$ when exposed to pure $CO_2$ and $CH_4$, respectively.

*Table 1. Boundary conditions used for finite element modeling of ZIF-8/SAW devices.*

| Boundary | Mechanical | Electrical |
|---|---|---|
| $\Gamma_L$, $\Gamma_{L1}$, $\Gamma_R$, $\Gamma_{R1}$ | Periodic | Periodic |
| $\Gamma_S$ | Free | Zero charge /symmetry |
| $\Gamma_G$ | Fixed | Ground |

## 3. Experimental

### 3.1 Sensor fabrication and characterization

SAW reflective delay lines of nominal center frequency 436 MHz were designed and fabricated in house on Y-Z $LiNbO_3$ piezoelectric crystal (*Roditi International*) by patterning 100 nm thick aluminum interdigitated electrodes (IDTs) using photolithography followed by lift off. The delay lines contained a bidirectional transmitting IDT ($T_r$) and two reflecting IDTs ($R_1$ and $R_2$) such that the reflectors were on opposite sides of the $T_r$ at 2.38-mm and 3.08-mm, respectively in the SAW propagating path. These distances were chosen to isolate the SAW reflections from the environmental echoes and to minimize the cross interferences. The IDTs $T_r$, $R_1$, and $R_2$ contained 50, 30, and 50 finger pairs, respectively of 2 μm wide electrodes separated by the same distance (i.e. 2 μm). The acoustic aperture of the IDTs were 100λ where λ is the SAW wavelength set by the IDT pitch (λ = 8 μm). The delay line in between $T_r$ and $R_2$ was coated with ZIF-8 whereas the one in between $T_r$ and $R_1$ remained uncoated and used as a reference delay line to account the environmental effects including changes in ambient temperature. For QCM measurements, the gold-coated AT-cut quartz crystal resonators (resonance frequency $f_0$ = 9 MHz, crystal diameter 25.4 mm, and piezoelectrically active area = 34.19 $mm^2$) and the measuring assembly that contained an oscillator circuit were purchased from INFICON.

ZIF-8 films of desired thickness were coated on all sides of the SAW and QCM substrates by a simple solution-based method at room temperature. First, the SAW delay lines were sonicated in acetone, ethanol, and DI water (five minutes each) and the QCM crystals were cleaned by piranha solution (*Caution! Piranha solution is acutely toxic chemical and is dangerous to handle without appropriate personal protective equipment and laboratory safety procedures*). The devices were rinsed into methanol and dipped into a homogenous solution of equal volume of zinc and 2-methylimidazole precursors in methanol solution with the active faces inclined down slightly. The reference delay line, the IDT $T_r$ and the reflector $R_1$ were tape-masked to avoid the deposition of the sensing material. The devices were taken out after thirty minutes, rinsed with methanol, and dried in $N_2$ gas that completed one cycle. Thicker films were obtained by repeating this cycle.

The morphology and thickness of the films were analyzed by scanning electron microscopy (SEM) (FEI QUANTA 600F). The crystal structures of the films were confirmed by X-Ray diffraction (XRD) (Panalytical X'Pert Pro diffractometer). The SAW devices were characterized in wired and wireless modes for delay and attenuation caused by the film using an R&S VZB vector network analyzer (VNA). The coated and uncoated QCM devices were tested for changes in resonance frequency and AC resistance due to the films using INFICON's RQCM.

Gravimetric gas adsorption measurements were conducted on commercial ZIF-8 (Basolite Z1200) using a Hiden IGA microbalance. The sample (~25 mg) was activated by heating under vacuum at 100°C until the sample weight was stabilized. Isotherms were then measured under flowing gases regulated by a mass flow controller. For measurements involving gas mixtures, the mixtures were prepared by blending gas flows in line prior to the sample cell using mass flow controllers set to the appropriate fractional flow rates. Equilibrium was determined at each pressure step using an internal fitting algorithm in the instrument control software. Buoyance corrections were applied to the final equilibrium weights using known densities of all components in the sample and counter weight chambers and gas densities calculated using REFPROP software from NIST. Data were collected out to a total pressure of 10 bar and the resulting isotherms were smoothed using fits to the Langmuir-Freundlich equation.

### 3.2 Gas sensing experiments

Sensing measurements were performed at room temperature and atmospheric pressure in real time in a gas cell consisting of a polyvinyl chloride tube (volume ~ 1 L) connected to an automated gas delivery system with a flow rate of 100 mL/min. The electric contacts to the SAW sensors were made by applying a conducting epoxy (MG chemicals) and curing at 80 °C overnight whereas those in QCM sensors were made using pogo pins. Then, the sensors were conditioned at least three hours in pure $N_2$ and then their responses to pure $CO_2$, $CH_4$, CO, $H_2$, and air followed by $N_2$ flush and to various concentrations of $CO_2/N_2$ and $CH_4/N_2$ mixtures were recorded in real time. The interrogator for SAW devices measurements was assembled as a custom system to record the phase difference of the emitting and reflecting waves in real time [4, 47]. The phase $\Delta\phi_{R1}$ of the reference delay line was subtracted from the phase $\Delta\phi_{R2}$ of the coated delay line ($\Delta\phi = \Delta\phi_{R2} - \omega\Delta\phi_{R1}$, where ω is the ratio of the distances of $R_2$ delay line to $R_1$ delay line from $T_r$) to eliminate environmental effects. The interrogator for QCM was based on an oscillator circuit that recorded the resonance frequency in real time and was commercially available (RQCM from INFICON). All gas testing experiments were done at room temperature and one atmospheric pressure in wired mode unless stated otherwise.

## 4. Results and Discussion

The sensing overlayer properties such as surface roughness, thickness, and interfacial mechanical adhesion with the piezoelectric substrate highly influence the performance of the acoustic sensors [2]. In the current study, growth of uniform ZIF-8 films with controlled thickness and smooth surface was achieved by directly applying its precursors on the surface [48]. This in-situ coating method provides much stronger mechanical adhesion than spin coating or drop casting of previously prepared solutions. To inspect the surface coverage and thickness of the grown films and to confirm the formation of ZIF-8 structure, we performed SEM imaging and XRD of the films coated on the transducers. **Figure 3(a)** is a representative SEM image of 3-cycle ZIF-8 film on Y-Z $LiNbO_3$ and **Figure 3(b)** and (**c)** are its magnified and cross-sectional images, respectively. The images depict that the grown films were dense, contained granular morphology, and fully covered the substrate. The grains had random shapes with the size (longer dimension) varying in between 100 – 200 nm for majority of them even though there were a few grains having long dimension as small as 40 nm and as large as 400 nm. The cross-sectional SEM image (**Figure 3(c)**) shows the growth of a uniform and continuous thin film with the thickness of about 300 nm with 3-cycles. Using similar deposition procedure, other researchers have also observed a growth of ~100 nm thick (which is equivalent to 95 ng/mm$^2$ mass, given the density is 0.95 g/cm$^3$) ZIF-8 films per cycle on different substrates [19]. In a closer look (inset of **Figure 3(c**)), we observed that the crystals had a columnar growth that indicated the nucleation of ZIF-8 crystals in each cycle [31]. The full coverage, homogeneous morphology, and uniform thickness of the film are important characteristics required for acoustic waves to propagate with low attenuation. The XRD data (**Figure 3(d**)) showed a sharp (001) peak at $2\theta = 7.5$ degree followed by other peaks (002), (112), and (222) at $2\theta$ = 10.5, 13, and 18.3 degrees, respectively. These peaks are in good agreement with the ZIF-8 XRD spectra reported in the literature[31] and confirmed the growth of ZIF-8 crystals on the substrate.

The fabricated SAW devices were examined by measuring their $S_{11}$ parameter and admittance over a frequency range of 415 MHz – 445 MHz for an input power of 0 dBm. **Figures**

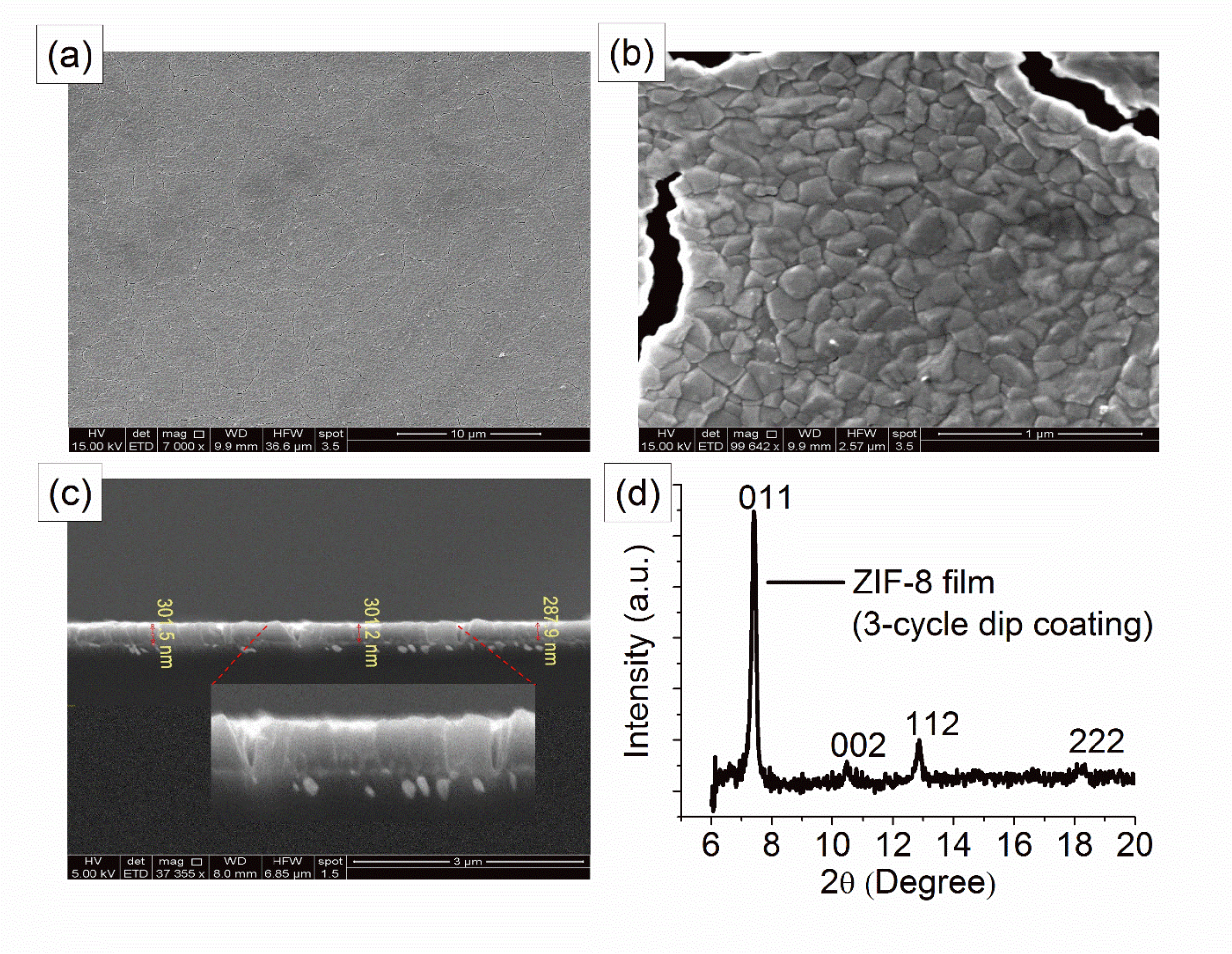

*Figure 3. Scanning electron microscopy (SEM) images (a-c) and X-Ray diffractometry (XRD) spectra (d) of 3-cycle dip-coated ZIF-8 films on Y-Z $LiNbO_3$. The cross-sectional SEM showed that 1-cycle dip coating of the precursors produced a 100-nm thick ZIF-8 film.*

**4(a)** and **4(b)** show typical $S_{11}$ magnitude responses of a SAW device before and after ZIF-8 coating (2 cycles) in time domain measured in wired and wireless modes, respectively. From the figure, one can see that the pulse responses of the device decreased rapidly until the acoustic reflections from $R_1$ and $R_2$ were observed. For this particular uncoated device, the $R_1$ and $R_2$ peaks were observed at 1.3576 μs and 1.7247 μs, respectively. Weak reflections observed before and after the acoustic reflections could be due to electromagnetic and acoustic reflections by the transducer edges. The $S_{11}$ spectra also contained the secondary reflections from $R_1$ and $R_2$. Similar features were observed from the coated device as well except that the acoustic waves travelling the coated channel (i.e. reflections from $R_2$) had a time delay and an attenuation while the position of the reflection peak from uncoated delay line $R_1$ did not have any observable change as expected. As seen in the figure, a weak peak was observed in between the reflections from $R_1$ and $R_2$ in the coated devices. This can be attributed to the reflections from the ZIF-8 overlayer. The time delay and reflection loss induced by 2-cycle coated (~200 nm) films were measured for five devices and employed to calculate the changes in SAW velocity and attenuation. The velocity and amplitude of the waves were found to decrease by 0.96% and $-3.08 \times 10^{-3}$ Np/rad, respectively due to the film. The measured velocity changes due to ~200 nm thick ZIF-8 film was approximately a factor of two larger than the calculated values using FEM. The discrepancy between the predicted value from the simulation and experimental observation could be due to the material properties assumed in the calculations which may differ from the actual values of the grown ZIF-8 films, presence of solvent molecules in the ZIF-8 pores, and uncertainties in the position of the reflection peaks introduced by the broad width amongst other potential factors. The effects of the ZIF-8 films coated by the same number of dip-coating cycles to the resonance frequency ($f_0$ = 9 MHz) of AT-Quartz resonators (i.e. QCM) were also studied. The films caused a decrease in the resonance frequency by ~4250 Hz, which corresponded to a mass density of about 1.17 g/cm$^3$, a slightly larger value than the one assumed in the simulation. It is important to note that both SAW and QCM observations showed a decrease in acoustic wave velocity when the transducers were coated with ZIF-8 films as predicted by the FEM simulation of SAW devices even though there are some discrepancies quantitatively. Potential reasons for the increased attenuation of SAWs in the coated devices are viscosity and surface roughness of the films [49]. However, it is important to note that the attenuation in the coated lines was also small enough to reflect the acoustic signals with a large and measurable intensity which was due to good rigidity, high compactness, and the smooth morphology of the films as well as their good interfacial mechanical adhesion with the piezoelectric substrate.

The coated SAW devices were examined wirelessly using a custom-fabricated and a commercial (SS433, LPRS) 433 MHz quarter wave whip antennae. The electric contacts of the devices with the antennae were made of silver conducting epoxy (purchased from MG Chemicals) and cured overnight at 80°C. The $S_{11}$ response shown in **Figure 4(b)** of the ZIF-8 coated device was measured using custom fabricated whip antennas placed at a separation of 4 cm. The figure shows that the reflection peaks were at expected positions but their intensities were reduced by ~20 dB. Much stronger peaks were observed with the commercial antenna (SS433, LPRS) at the same distance and the interrogation was possible from a distance as far as 45 cm. Therefore, the gas sensing test with wireless interrogation to be explained below was performed using the commercial antennas.

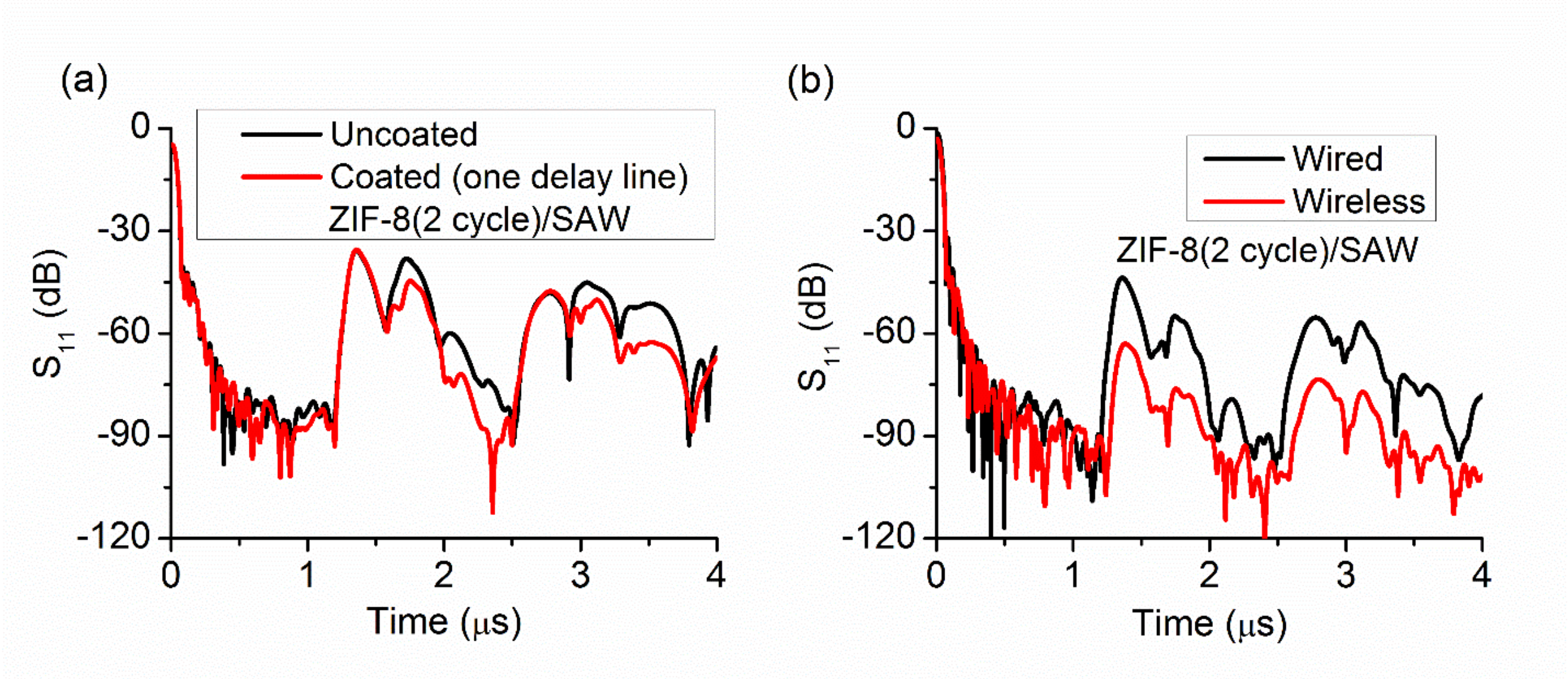

*Figure 4. Electric characterization of uncoated and coated SAW devices in (a) wired and (b) wireless modes. The coated devices consisted of a delay line coated with 2 cycles of ZIF-8 film and an uncoated reference delay line. Shown wireless measurements were performed.*

Gas adsorption isotherms were measured on bulk ZIF-8 powder to determine the relative mass changes which would be expected in a device using gas sorption in ZIF-8 for a sensing response. **Figures 5(a)** and **5(b)** show the gravimetric adsorption isotherms for $CO_2/N_2$ and $CH_4/N_2$ mixtures at 25 °C, respectively. At 1 bar, the measured adsorptions of pure $CO_2$, $CH_4$, and $N_2$ in ZIF-8 were 28.3 mg/cm$^3$, 4.11 mg/cm$^3$, and 2.41 mg/cm$^3$, respectively. The adsorption of these gases in ZIF-8 occurs through a physisorption mechanism and is completely reversible. The magnitude of the gravimetric response is dependent on two main factors. Physisorption depends on the polarizability of the gas to provide van der Waals interactions with the pore surface and the compatibility of the kinetic diameters of the gases with the aperture size of MOF pores. Therefore, the molar adsorption potential of a series of non-polar gases correlates well with their critical temperatures such that $CO_2 > CH_4 > N_2$ [50]. However, since the actual property being measured is the mass change of the sorbent upon gas adsorption, the molecular weight of the gas has a significant effect on the final value. When the effects of the molar adsorption potentials and molecular weights of the gases are combined, the gravimetric response for $CO_2$ becomes much larger than that observed for $CH_4$ and $N_2$. In addition, when comparing $CH_4$ and $N_2$ adsorption, the larger molecular weight of $N_2$ partially compensates for its lower molar adsorption potential to yield a gravimetric response which is only slightly below that of $CH_4$. The effects of these combined influences can be seen in the mixed gas isotherms of **Figure 5**. For $CO_2/N_2$ mixtures, the high adsorption affinity and high molecular weight of $CO_2$ give a large difference in adsorbed mass as composition is varied between pure $N_2$ and pure $CO_2$. For mixtures of $CH_4/N_2$, the low molecular weight of $CH_4$ yields a much smaller difference in adsorbed mass as the gas composition is varied between pure $N_2$ and pure $CH_4$. It is clear from the plots in Figure 5 that a mass based sensor using ZIF-8 will show a mass response to changes in $CH_4$ or $CO_2$ compositions in a $N_2$ background and the sensor will have a significantly higher sensitivity to $CO_2$ than to $CH_4$. In effect, a calibration curve for the SAW device can be constructed by plotting the total mass of the gas adsorbed at 1 bar for each gas mixture versus the volume percent of the respective gas in the $N_2$ background. The results are the linear plots shown in **Figures 5(c)** and **5(d)**. These plots show that any $CO_2/N_2$ or $CH_4/N_2$ mixture will give a unique value of total mass adsorbed in the ZIF and demonstrate the potential utility of such a device for monitoring emissions of $CO_2$ or $CH_4$ in air. The ratio of the sensitivity of $CO_2$:$CH_4$ for the ZIF-8 sensor can be estimated to be approximately 17:1 from the ratio of the slopes of the $CO_2/N_2$ and $CH_4/N_2$ plots in **Figures 5(c)** and **5(d)**. Adsorption of different amount of these gases into the material through physisorption has direct effect on the sensor responses to be discussed below.

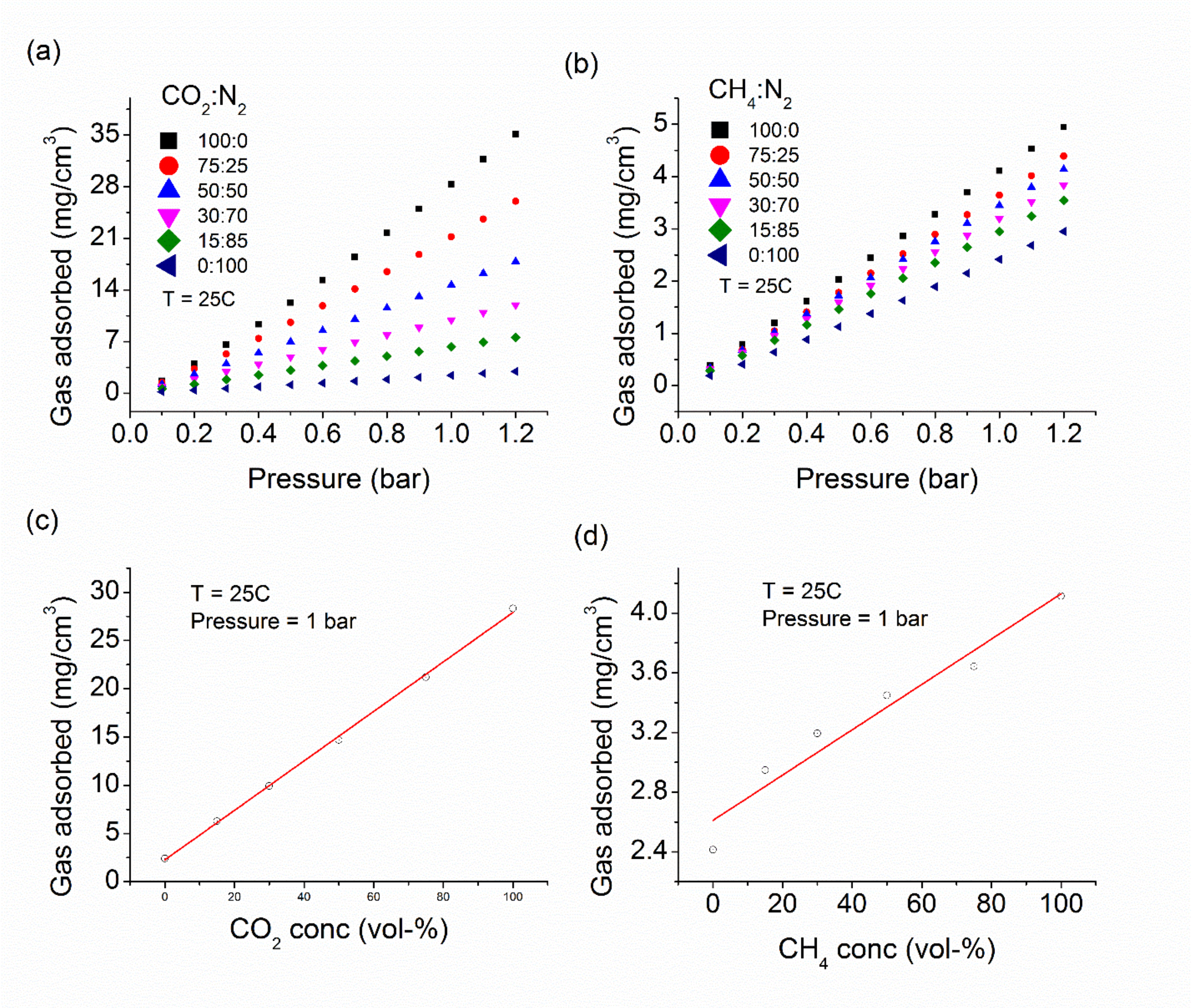

*Figure 5. Gas adsorption isotherms for $CO_2/N_2$ (a) and $CH_4/ N_2$ (b) mixtures in ZIF-8 powder measured by gas chromatography at 25°C. (c) and (d) are the total $CO_2$ and $CH_4$ adsorbed in the powder at 1 bar and 25°C for various compositions of $CO_2/N_2$ and $CH_4/N_2$ mixtures, respectively.*

The responses of SAW reflective delay lines to the exposed gases can be measured in terms of time delay or phase in wired (**Figure 6 (a)**) or wireless (**Figure 6(b)**) modes. In our experiments, radio frequency (RF) pulses were excited at frequency $f_0$ where sharp and clearly resolved reflection peaks were observed. Then the phase shift of the reflected signal with reference to the phase of the emitted signal was recorded in real time as it provides higher sensitivity and resolution than the time delay. **Figure 6(a)** is a typical real-time phase response of a 3-cycle coated SAW sensor ($f_0$ = 431.4 MHz) measured in wired mode when exposed to various gases ($CO_2$, $CH_4$, $H_2$, CO, and air) followed by $N_2$ flush. When the test gas began flowing through the cell, the phase of the sensor began to shift from the original value ($N_2$ baseline) and achieved a saturation in ~12 minutes. The sensor had a significant and distinct phase change for pure $CO_2$, $CH_4$, and $H_2$, negligible change for pure CO, and no observable change for air relative to the $N_2$ baseline phase. The phase of the sensor was shifted by -1.396 rad for $CO_2$, 0.136 rad for $H_2$, -0.073 rad for $CH_4$. In case of $H_2$, however, the phase shift was in the opposite direction relative to the shift for other gases. When the cell was flushed by $N_2$, the phase returned to the original $N_2$ baseline quickly demonstrating a good reversibility of the sensor. A detailed analysis of the response and recovery times are not presented here because the effective response time is dominated by the relatively

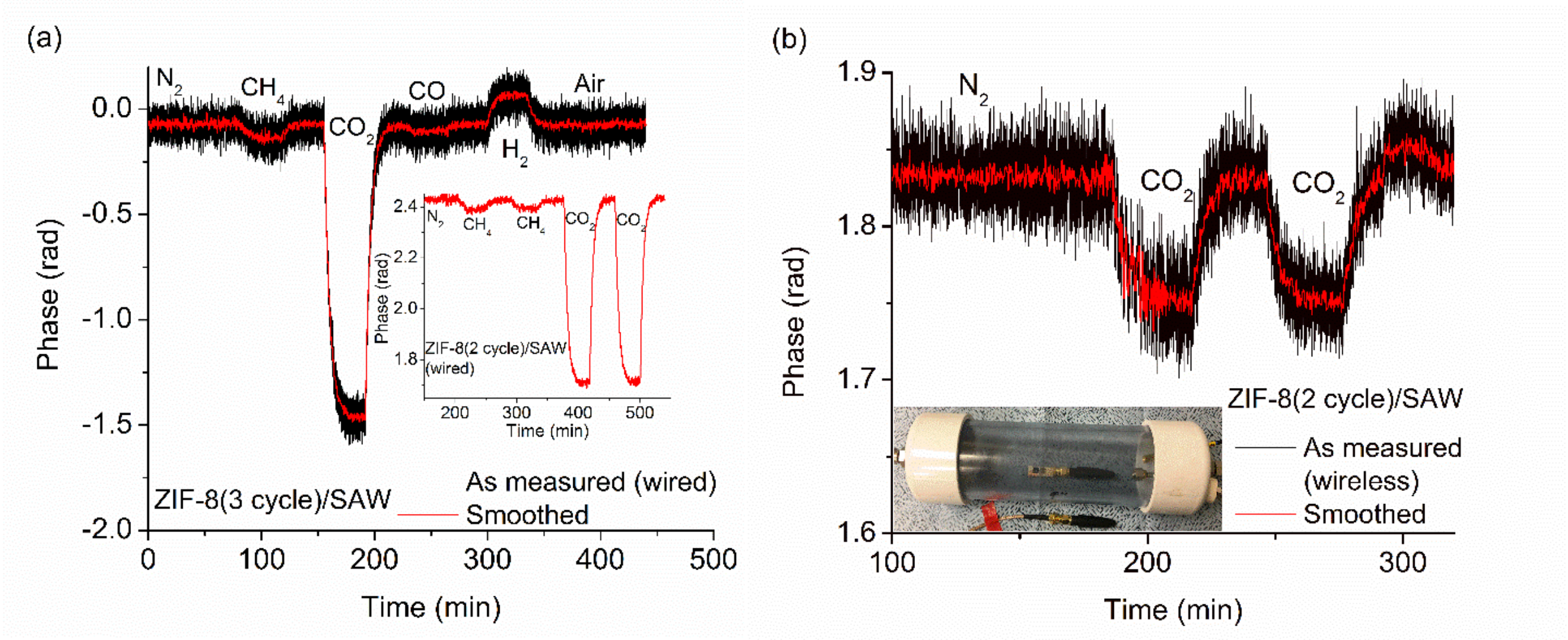

*Figure 6 (a) Real time phase of a 3-cycle ZIF-8 film-coated SAW sensor measured in wired mode for various gases exposure. Inset is the response of a 2-cycle film-coated SAW sensor to pure $CO_2$ and $CH_4$ for repeated exposure. (b) Wirelessly measured real time phase of a 2-cycle film-coated SAW sensor for pure CO2 (70 L/min). Shown wireless measurement was performed using commercial whip antennas.*

large volume of the reactor utilized in the experiments in question. When the sensor was mounted into a smaller gas cell, the sensor responded much faster which was consistent with the previously demonstrated response times (~ 10 sec) for the MOF films of similar thickness integrated with an optical fiber platform [48]. Inset of **Figure 6(a)** shows a typical phase value of a 2-cycles coated SAW sensor ($f_0$= 431.4 MHz) in real-time for two cycle flow of pure $CO_2$ and $CH_4$. As shown, the sensor had excellent reversibility and repeatability to both gases. In this case, the phase was shifted by -0.72 rad and -0.04 rad for pure $CO_2$ and $CH_4$, respectively, relative to pure $N_2$ base line. The observed shifts corresponded to the fractional velocity (*i.e.* phase, see Eq. – 2) changes of $152.66\times10^{-6}$ and $8.4811\times10^{-6}$ for pure $CO_2$ and $CH_4$, respectively considering the film covering the full delay path and causing a delay of $t_0$ = 1.7400 µs when gases are not exposed. **Figure 6(b)** shows the response of a 2-cycles ZIF-8 coated SAW device ($f_0$= 430 MHz) to pure $CO_2$ (70 L/min) as measured wirelessly using commercial whip antennas (shown in the inset of the figure). The observed excellent sensitivity with good reversibility and repeatability demonstrated the wireless capacity of the sensor. Note that, all measurements other than this demonstration were performed in wired mode. Excellent reversibility of the sensor is the signature of the physisorption-based absorption. Similarly, the repeatable response indicated a good mechanical interfacial adhesion between the sensing layer and the piezoelectric substrate and the stability of the device in ambient conditions.

The phase change upon exposure to various gases relative to the phase for $N_2$ is attributed to the change the net mass of sensing film (Eq. – 1) due to the difference in the material's adsorption capacity to different gases as well as variation in the molecular weight of the gases. For $CO_2$ and $CH_4$, the net mass of the film increased compared to the net mass with $N_2$ so that the velocity decreased and hence the phase shifted in the negative direction. Similar to the observation of the $CO_2$ adsorption isotherms in ZIF-8 powder shown in **Figure 5(a**), ZIF-8 films on the sensor adsorbed large amounts of heavier $CO_2$ molecules so that its effective mass increased significantly causing a large phase shift. Similarly, there was some increase in the net mass of the film with exposure to $CH_4$ even though $CH_4$ molecules are lighter than $N_2$ since the ZIF-8 films adsorbed greater amounts of $CH_4$ due to its higher adsorption potential relative to $N_2$. Consequently, the phase shifted in the same direction as for the case of $CO_2$ but with much smaller magnitude which agreed with the gas adsorption isotherms shown in **Figure 5**. Since the adsorption potential and molecular weights of air ($N_2/O_2$), $N_2$, and CO are all very similar, the sensor did not easily detect changes in gas composition involving these gases. The reverse phase shift for $H_2$ was expected since $H_2$ has a much lower adsorption potential and lower molecular weight than $N_2$. Thus, as the volume percent of $H_2$ increased, the amount of $N_2$ adsorbed in the ZIF decreased resulting in an overall mass loss in the ZIF sensing layer. The phase shift for pure $CO_2$ was 18 times larger than that for pure $CH_4$. This ratio is commensurate with the 17:1 adsorbed mass ratio of $CO_2$ to $CH_4$ at 1 bar observed in the gravimetric isotherm measurements for bulk ZIF-8. This correlation confirms the structural integrity of the ZIF-8 SAW coating and verifies that the physical properties of the bulk material are precisely incorporated into the sensor response. Compared to FEM calculations, slightly smaller values of the fractional velocity changes of the fabricated 200 nm ZIF-8 SAW devices were observed for the pure gases. A potential reason for this difference could be the partition coefficients used in the FEM calculations. The dimensionless Henry's constants evaluated from the adsorption isotherms (Figure 5(a) and 5(b)) for pure $CO_2$ and $CH_4$ were used as the partition coefficients in the calculation whereas the sensor responses were recorded in $N_2$ background at 1 atm. Second, the ZIF-8 films were considered rigid in FEM calculation so that the changes in mechanical properties, if any, upon gas adsorption were ignored. Other reasons for the deviated response could be the consideration of the full film coverage of the delay path and the error in the estimation of the reference delay time.

**Figures 7(a)** and **8(b)** depict the typical dynamic phase responses of the 2-cycles film coated SAW sensor to various concentrations of $CO_2$ and $CH_4$, respectively. As expected, the sensor showed smaller phase change when $CO_2$ ($CH_4$) volume was decreased in the $CO_2/N_2$ ($CH_2/N_2$) mixture due to a decrease in net mass change of the film. The maximum phase shifts were extracted from the dynamic response and plotted against the gas concentrations in **Figures 7(c) and 7(d**). The responses were fit linearly, the slopes of which were evaluated as the sensor's sensitivity to the respective gases. The evaluated sensitivities were 0.394 deg/vol-% and 0.021 deg/vol-% for $CO_2$ and $CH_4$, respectively. We also evaluated the sensor's sensitivity in terms of (fractional change in signal)/concentration, *i.e.* $\eta_{SAW} = (\Delta v/v_0)/\mathrm{C_{gas}}$ (where $\mathrm{C_{gas}}$ is the volume fraction of the gases in $N_2$) that yielded the sensitivities of $1.44\times10^{-6}$/vol-% and $8\times10^{-8}$/vol-% for $CO_2$ and $CH_4$, respectively.

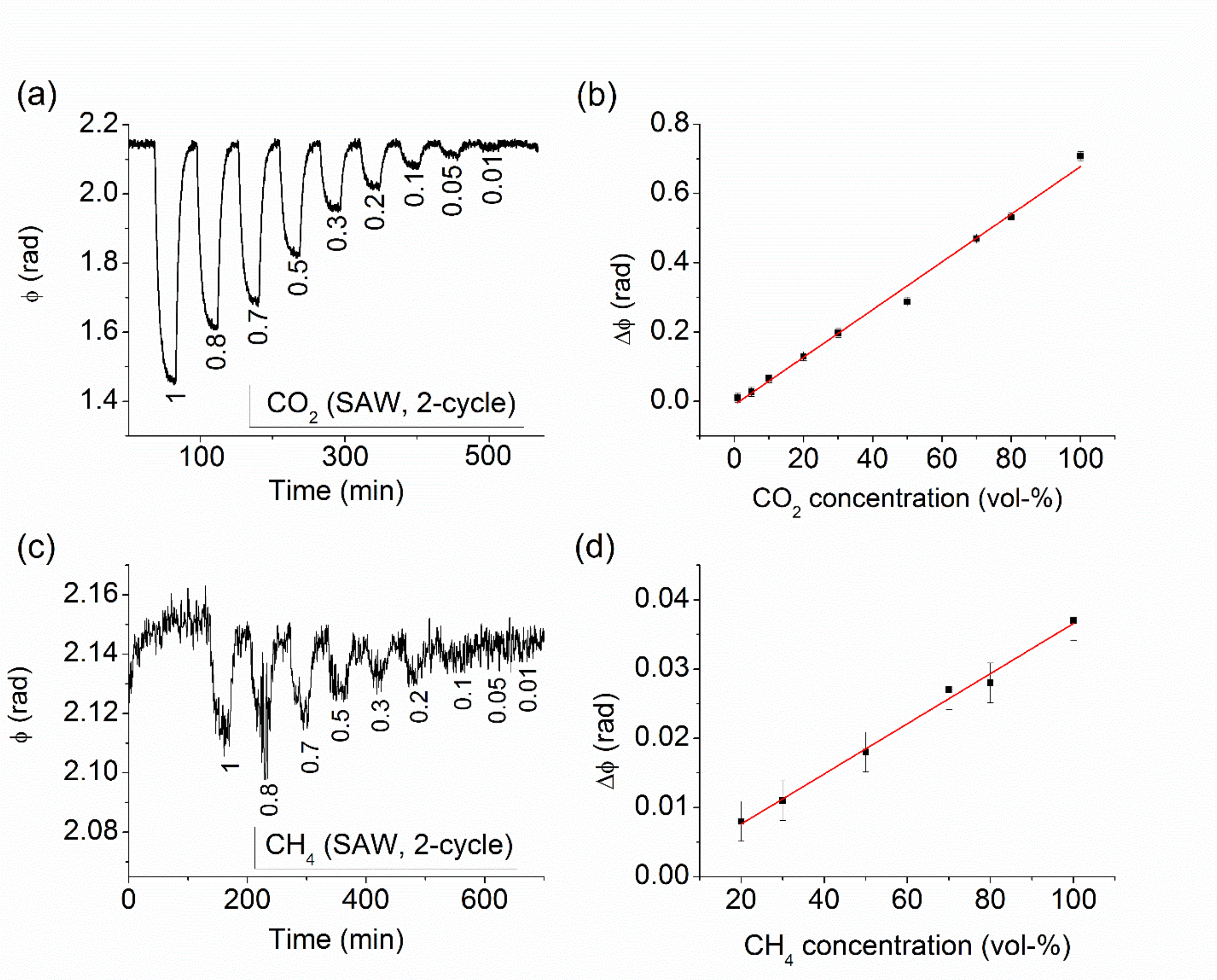

*Figure 7. Response of a SAW sensor coated with 2-cycle ZIF-8 film to various concentrations of $CO_2$ and $CH_4$ in $N_2$.*

**Figures 8(a)** and **8(c)** show the real-time frequency responses of a 2-cycles coated 9 MHz QCM sensor to various concentrations of $CO_2$ and $CH_4$ in $N_2$. As expected, the frequency started decreasing when the sensor was exposed to the gases, reached a minimum and saturated. When the gases were flushed away by pure $N_2$, the frequency returned close to the original base line of $N_2$ indicating a minimal hysteresis of the sensor. It is important to note that reference data were not collected during the gas testing by QCM unlike in the case of SAW measurements. For pure $CO_2$ and $CH_4$, the maximum decrease in the frequency was 217 Hz (fractional change: $24.11\times10^{-6}$) and 9.2 Hz (fractional change: $1.02\times10^{-6}$), respectively giving to a ratio of ~ 23.6. **Figures 8(b)** and **8(d)** show the maximum frequency shift extracted from the real-time data for various concentrations of $CO_2$ and $CH_4$. The shifts in frequency varied linearly with the concentrations giving rise to the sensitivities (obtained as the slope of the liner fits) of 2.18 Hz/vol% (~ $0.24\times10^{-6}$/vol-%) and 0.09 Hz/vol-% ($1\times10^{-8}$/vol-%) for $CO_2$ and $CH_4$, respectively.

The decrease in the frequency for $CO_2$ and $CH_4$ relative to $N_2$ baseline is attributed to the increased mass density of sensing film upon the gas exposure as explained for the case of the SAW sensor. Since the $CO_2$ molecules are much heavier and small enough to be incorporated into the ZIF-8 cages, the QCM frequency shift was much larger than for $CH_4$, consistent with the SAW output and adsorption isotherms. The QCM sensor sensitivity for $CO_2$ was about 23.6 times higher than that for $CH_4$ whereas the SAW sensor sensitivity for $CO_2$ was about 18 times higher than that for $CH_4$ as obtained from the slopes of the sensors outputs versus concentration. These values are comparable to the $CO_2$ to $CH_4$ mass adsorption ratio of ~17 determined for ZIF-8 powder (**Figures 5(c)** and **5(d))**. The slopes of the linear responses of the sensors were converted into the mass adsorbed per unit volume of ZIF-8 film (2-cycles ZIF-8 coated) for one vol-% exposure of the gases in $N_2$ environment using Eq. 1 – 3. The evaluated amounts of adsorbed $CO_2$ and $CH_4$ per unit volume of ZIF-8 films were close to the corresponding mass adsorbed in unit volume of ZIF-8 powder obtained by gravimetric measurement which are summarized in **Table 2**.

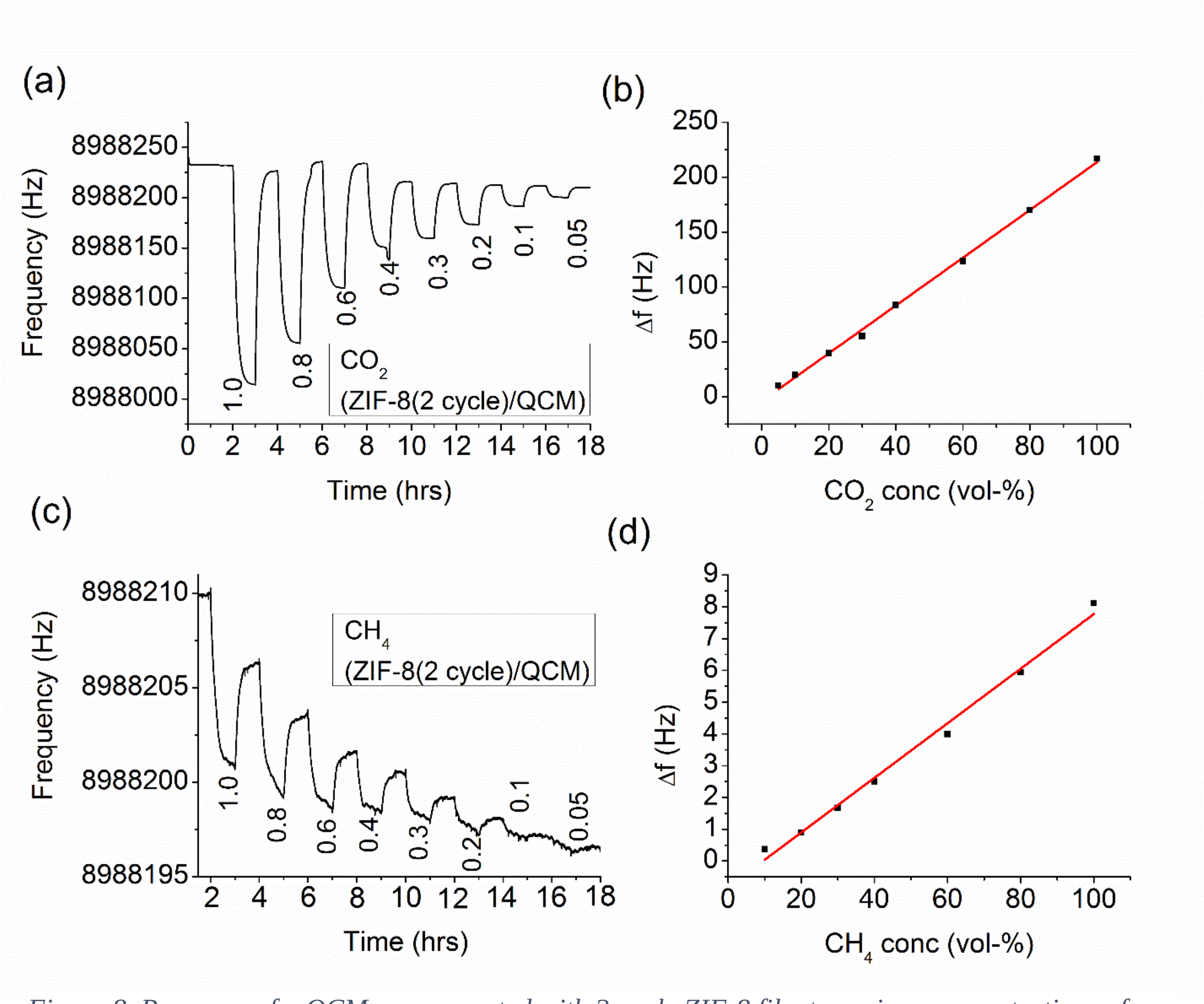

*Figure 8. Response of a QCM sensor coated with 2-cycle ZIF-8 film to various concentrations of $CO_2$ and $CH_4$ in $N_2$.*

*Table 2. Amount of gas adsorbed into ZIF-8 films obtained from SAW and QCM sensors and in to ZIF-8 powder obtained from gravimetric method. The mass adsorbed per unit volume of the films for exposure of unit vol-% of the gases in N2 were calculated by Eq. 1 – 3 using the slope of the sensor output versus concentration.*

| Gas ($mg/cm^3$)/vol-% | Gravimetric (powder) | QCM | SAW |
|---|---|---|---|
| $CO_2$ ($mg/cm^3$)/vol-% | 0.257 | 0.595 | 0.304 |
| $CH_4$ ($mg/cm^3$)/vol-% | 0.015 | 0.025 | 0.016 |

Even though the ratio of the sensor outputs to the gases were comparable, the actual fractional SAW sensitivities were higher than the fractional QCM sensitivities to either gas. Direct comparison of the sensor sensitivities to a given gas is difficult as the sensors were fabricated in different piezoelectric substrates and the exposed area for SAW sensors were much smaller than that for QCM sensors. However, it is worth to note that the SAW transducers generally yield a higher sensitivity than QCM due to their surface-confined propagation mode and higher frequency of operation [51]. As the SAWs propagate along the surface, their acoustic energy is mostly confined to the sheath of the surface so that they suffer stronger interaction with the environment unlike the bulk waves that propagate across the piezoelectric crystal in thickness shear mode. Also, the mass loading sensitivity of acoustic sensors is proportional to the square of operational frequency (Eq. 1 and 3), that makes the SAW sensors able to achieve much higher sensitivity compared to QCM sensors given that the mass sensitivity coefficient are similar [41]. Potential reasons for deviations of the adsorbed mass in the films calculated by the sensor outputs from that obtained by the gravimetric measurements in powder could be the due to the film's viscoelastic contribution or variations in the interfacial mechanical adhesion with the substrates amongst others [38, 52]. Further investigation is needed to understand the mechanical changes of the ZIF-8 films upon gas adsorption and the influence of such changes to the sensor output [53, 54].

To understand the effect of the ZIF-8 film thickness on the SAW response, we measured the responses of 1, 2, and 3 cycles coated sensors to various concentrations of $CO_2$ and $CH_4$, the results of which are shown in **Figure 9**. As shown, the sensitivity of the sensor increased with increasing the number of coating cycles *i.e.* the thickness. The calculated sensitivities of the 1, 2, and 3 cycles ZIF-8 coated devices were 0.138 deg/vol-%, 0.394 deg/vol-%, and 1.038 deg/vol-%, respectively for $CO_2$ and 0.006 deg/vol-%, 0.021 deg/vol-%, and 0.042 deg/vol-%, respectively for $CH_4$. With increasing the thickness, the film provided more sites for gas molecules to be incorporated. Consecutively, the net areal mass density of the film increased resulting to a higher sensitivity of the sensor. From the measured data, we observed a linear increase in the sensitivity for the film thickness increasing from 100 nm to 300 nm. However, the sensitivity is not expected to retain a linear increase with film thickness indefinitely [2] for several reasons. First, kinetic limitations will eventually lead to an unacceptably long time constant due to the need for diffusion through the entire depth of a thick film. Second, the response is known to become significantly more complex in the case of acoustically thick films. When the film is thick, not all of the acoustic energy is stored in the substrate and the velocity shift is no longer exactly proportional to the film mass [38]. In addition, MOF films have small transverse velocity leading to standing waves in thicker films. There are also chances of the changes in the film morphology and crystal orientation with increased thickness [31] to affect the uptake of gases, which potentially affect the sensor sensitivity.

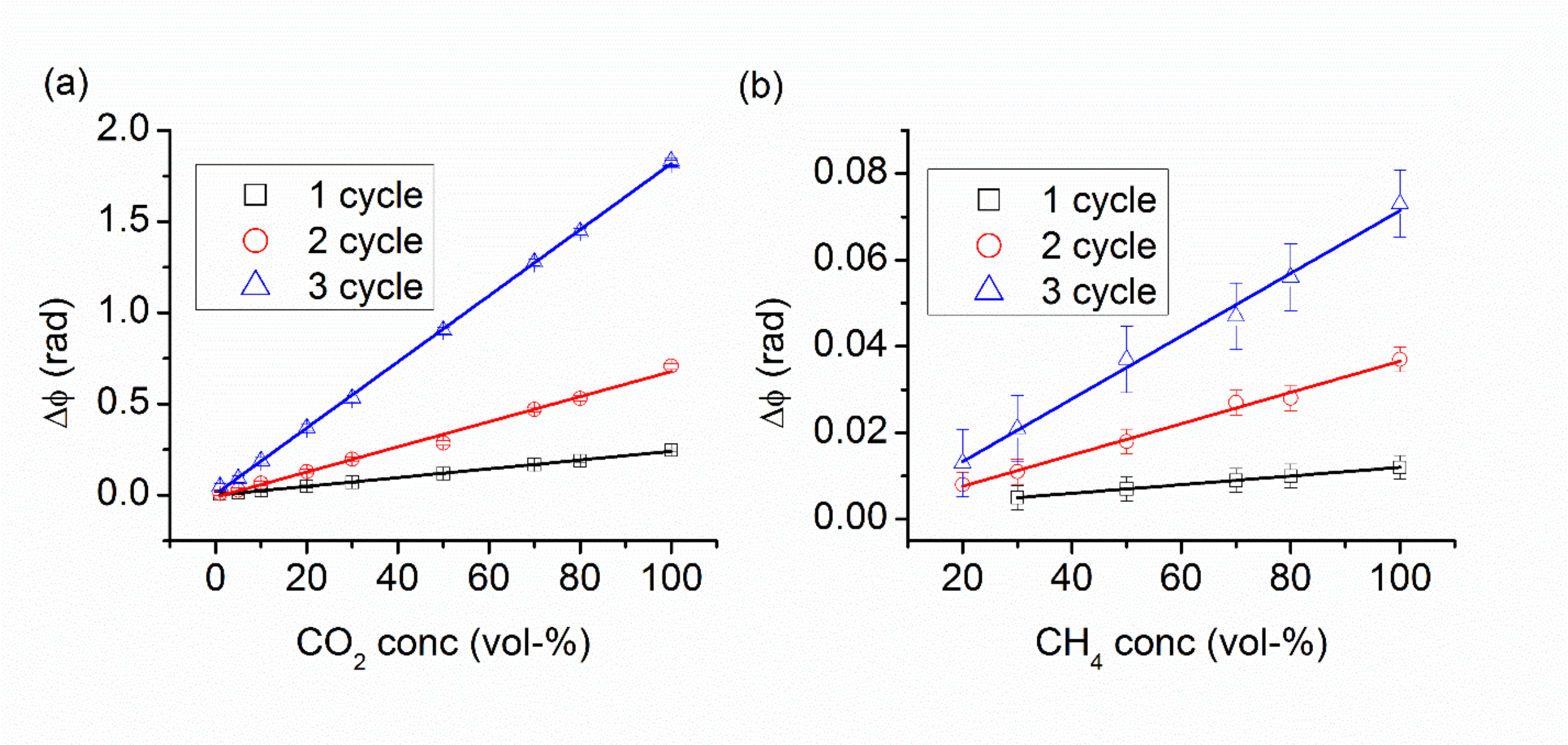

*Figure 9. Responses of SAW sensors with varied ZIF-8 film thickness (100 nm – 300 nm) to various concentrations of (a) $CO_2$ (a) (b) $CH_4$.*

## 5. Conclusion

Surface and bulk acoustic wave sensors were successfully developed using nanoporous ZIF-8 MOF as the sensing overlayer for monitoring $CO_2$ and $CH_4$ in $N_2$ at ambient temperature and pressure. The SAW sensors' sensitivity increased with the thickness of the overlayer over the range 100 nm – 300 nm, which was attributed to the adsorption of larger amount of gases into the thicker films. The SAW sensors were found to possess higher sensitivity (in terms of fractional change in the signal) than the QCM sensors to the gases tested for similar overlayer thicknesses. In addition, the coated SAW devices had good acoustic reflections when operated in wireless mode indicating its potential wireless and passive application. The SAW sensors showed reversible and reproducible responses relative to a $N_2$ baseline for incremental additions of $CO_2$ or $CH_4$ to the N2 sweep gas. In contrast, no measurable response was observed when CO or air was added to the sweep gas. The sensors showed an enhanced sensitivity to $CO_2$ relative to $CH_4$ due to the higher adsorption potential and higher density of $CO_2$. The sensor responses for $CO_2/N_2$ and $CH_4/N_2$ mixtures correlated well with the mass gains determined for bulk ZIF-8 from gravimetric isotherms measurements and thus confirmed that the physical properties of the bulk ZIF-8 are preserved in the sensor coating. By correlating the linear response of the sensor to the volume percent of $CO_2$ or $CH_4$ in $N_2$, a simple calibration curve can be constructed which provides a unique mass reading for each point along the concentration gradient of the $CO_2/N_2$ or $CH_4/N_2$ mixtures and demonstrates the potential of these devices for monitoring $CO_2$ or $CH_4$ in environmental and emission sites across the energy infrastructure including pipelines, active and abandoned wells, compressors, and others. Considering the proven tailorability of gas adsorption properties for the thousands of reported and hypothetical MOFs in the scientific literature, the current demonstration of MOF-coated SAW gas sensors is of general importance in developing passive gas sensors for targeted applications with the potential for remote and wireless interrogation.

**Acknowledgement**

This technical effort was performed in support of the National Energy Technology Laboratory's ongoing research in Natural Gas Transmission and Delivery FWP Support under the RES contract DE-FE0004000.

**Disclaimer**

This project was funded by the Department of Energy, National Energy Technology Laboratory, an agency of the United States Government, through a support contract with AECOM. Neither the United States Government nor any agency thereof, nor any of their employees, nor AECOM, nor any of their employees, makes any warranty, expressed or implied, or assumes any legal liability or responsibility for the accuracy, completeness, or usefulness of any information, apparatus, product, or process disclosed, or represents that its use would not infringe privately owned rights. Reference herein to any specific commercial product, process, or service by trade name, trademark, manufacturer, or otherwise, does not necessarily constitute or imply its endorsement, recommendation, or favoring by the United States Government or any agency thereof. The views and opinions of authors expressed herein do not necessarily state or reflect those of the United States Government or any agency thereof.